\documentclass{elsart4-1}

\usepackage{amssymb}

\usepackage[english,francais]{babel}

\newtheorem{e-proposition}[theorem]{Proposition}

\newtheorem{e-definition}[theorem]{Definition\rm}

\renewcommand{\theequation}{\arabic{equation}}
\def\og{\leavevmode\raise.3ex\hbox{$\scriptscriptstyle\langle\!\langle$~}}
\def\fg{\leavevmode\raise.3ex\hbox{~$\!\scriptscriptstyle\,\rangle\!\rangle$}}

\usepackage{graphicx}

\begin{document}

\centerline{Physics or Astrophysics/Header}
\begin{frontmatter}

\selectlanguage{english}
\title{Cosmic Rays from the Ankle to the Cut-Off}

\selectlanguage{english}
\author[authorlabel1]{Karl-Heinz Kampert},
\ead{kampert@uni-wuppertal.de}
\author[authorlabel2]{Peter Tinyakov}
\ead{petr.tiniakov@ulb.ac.be}

\address[authorlabel1]{University Wuppertal, Department of Physics, D-42119 Wuppertal, Germany}
\address[authorlabel2]{Service
 de Physique Th\'eorique, Universit\'e 
Libre de Bruxelles, CP225, Boulevard du Triomphe, B-1050 Brussels, Belgium}

\medskip
\begin{center}
{\small Received *****; accepted after revision +++++}
\end{center}

\begin{abstract}
Recent advances in measuring and interpreting cosmic rays from the
spectral ankle to the highest energies are briefly reviewed. The prime
question at the highest energies is about the origin of the flux
suppression observed at $E \simeq 4\cdot10^{19}$\,eV. Is this the long
awaited GZK-effect or the exhaustion of sources? The key to answering
this question will be provided by the largely unknown mass composition
at the highest energies. The high level of isotropy observed even at
the highest energies challenges models of a proton dominated
composition if extragalactic magnetic fields are on the order of a few
nG or less. We shall discuss the experimental and theoretical progress
in the field and the prospects for the next decade.

\vskip 0.5\baselineskip

\selectlanguage{francais}
\noindent{\bf R\'esum\'e}
\vskip 0.5\baselineskip
\noindent
{\bf Rayons cosmiques de la cheville \'{a} la coupure}
Les avanc\'{e}es r\'{e}centes concernant les mesures et l'interpr\'{e}tation des rayons cosmiques, de la caractŽristique spectrale appel\'{e} cheville jusqu'aux plus hautes \'{e}nergies, sont bri\'{e}vement revues. Aux plus hautes \'{e}nergie, la question principale concerne l'origine de la suppression du flux clairement observ\'{e}e au dessus de $4\times10^{19}$\,eV. Est-ce la pr\'{e}diction GZK tant attendue? Ou bien l'\'{e}puisement des sources? La clef qui permettra de r\'{e}pondre \`{a} cette question r\'{e}side dans la composition des rayons cosmiques qui est aujourd'hui largement inconnue aux plus hautes \'{e}nergies. L'isotropie observ\'{e}e dans la distribution des directions d'arriv\'{e}es, m\^{e}me aux plus hautes \'{e}nergies, d\'{e}favorise les mod\'{e}les o\'{u} les protons dominent la composition si les champs magn\'{e}tiques extragalactiques sont au niveau de quelques nG ou moins. Nous discuterons les progr\'{e}s exp\'{e}rimentaux et th\'{e}oriques du domaine et les perspectives pour la prochaine d\'{e}cennie.

 \vskip 0.5\baselineskip
\noindent{\small{\it Mots-cl\'es~:} Rayons cosmiques; Composition masse; Anisotropie}}
\end{abstract}
\end{frontmatter}


\selectlanguage{english}

\section{Introduction}
\label{sec:intro}

In the last decade, a new generation of the ultra-high energy cosmic ray
(UHECR) observatories has come into operation: the Pierre Auger Observatory in
the Southern hemisphere and the Telescope Array in the Northern one.  Apart
from a significant advance in size over their predecessors, both observatories
have implemented, for the first time, a new hybrid technique of the UHECR
detection where the same air shower is observed simultaneously by a ground
array of particle detectors and by fluorescence telescopes capable of tracing
the development of the air shower in the atmosphere. By now, both
observatories have accumulated a significant part of their lifetime
statistics. It may be time, therefore, to summarize the advances in our
understanding of UHECR and formulate the remaining problems.

The Pierre Auger Observatory (Auger) \cite{Abraham:2004dt} is located
in Argentina (centered at $69^\circ20$ W, $35^\circ20$ S) at 1400\,m
above sea level, corresponding to 870\,g/cm$^2$. It consists of a
Surface Detector array (SD) comprising 1660 autonomously operated
water-Cherenkov detectors of 10\,m$^2$ area each. The tanks are filled
with 12 tons of purified water and three photomultipliers are used to
detect the Cherenkov light produced by charged particles.  The surface
detectors are spread over 3000~km$^2$ area and are placed on a
triangular grid of 1.5 km spacing. The SD array is overlooked by 27
fluorescence detector telescopes (FD) distributed at five sites
\cite{Abraham:2009pm}. Stable data taking started in January 2004 and
the Observatory has been running with its full configuration since
2008.

The Telescope Array (TA) is located in Utah, USA at $39^\circ
30$~N, $112^\circ 91$~W at an altitude of about 1400~m above sea
level. It consists of 507 plastic scintillator detectors of
3~m$^2$ area each spread over approximately 700\,km$^2$ (for
details see \cite{AbuZayyad:2012kk}). The detectors are placed on
a square grid with a spacing of 1.2~km. The atmosphere over
the surface array is viewed by 38 fluorescence telescopes
arranged in 3 stations \cite{Tokuno:2012mi}. TA is fully
operational since March 2008.

Despite similar hybrid design, the two experiments have a number of
differences that should be kept in mind when comparing the results.
The main one is the design of the ground array detectors. The
detectors of TA are traditional two layers of 1.2~cm thick plastic
scintillators, similar to the single 5~cm thick layers used in AGASA.
The water tanks of the Pierre Auger Observatory have a thickness of
1.2~m and a large overall volume, that makes them more sensitive than
the TA detectors, especially to inclined particles. At the same time,
the large thickness enhances the signal due to the penetrating muonic
component of a shower, which is more difficult to model. 

By now, an unprecedented number of UHECR events have been detected by
the ground arrays and the fluorescent telescopes of both
experiments. At energies $E>10^{19}$~eV over $10^4$ events have been
recorded by the Pierre Auger Observatory, and over $2\times 10^3$ by
the Telescope Array. For each event, several observables can be reconstructed, the
key ones being the energy of the primary particle, the arrival direction
and, for the events detected by the fluorescence telescopes, the
atmospheric depth of the air shower maximum. These and other
observables allow one to shed some light on the nature of primary particles
and the origins of UHECR, as discussed in the next sections.

\section{Energy Spectra}
\label{sec:energy}

The all-particle energy spectrum is perhaps the most prominent
observable of cosmic
rays being investigated. It carries combined information about the
UHECR sources and about the galactic and/or intergalactic media in
which CRs propagate. The ankle, a hardening seen in the all-particle
spectrum at about $5\cdot10^{18}$\,eV, is generally considered to
mark the transition from galactic to extragalactic cosmic rays.
However, recent measurements of KASCADE-Grande
\cite{Apel:2011mi,Apel:2013ura} suggest that this transition may occur
more than an order of magnitude lower in energy, i.e.\ around
$10^{17}$\,eV. At this energy, the component of light elements is
subdominant but exhibits a hardening to become dominant at the ankle.
The so-called dip-model of the ankle \cite{Aloisio:2012ba} interprets
the ankle as being the imprint of protons suffering $e^+ e^-$
pair-production in the CMB. Thus, it requires protons to be dominant
at energies significantly above and below the ankle and the transition
to occur again below the ankle energy. Obviously, models differ in
their energy spectra expected for different mass groups and thereby in
their cosmic ray mass composition as a function of energy. Related to
this, one also expects to see different levels of anisotropies in the
arrival directions as it will be difficult to fully isotropize EeV
protons in galactic magnetic fields \cite{Ptuskin93}.

At the highest energy, a flux-suppression due to energy losses by
photo-pion production and photo-disintegration in the CMB is expected
for protons and nuclei, respectively. In fact, this so-called
GZK-effect \cite{Greisen:1966jv,Zatsepin66} is the only firm
prediction ever made concerning the shape of the UHECR spectrum. First
observations of a cut-off were reported by HiRes and Auger
\cite{Abbasi:2007sv,Abraham:2008ru}. However, at present we cannot be
sure whether this flux suppression is an imprint of the aforementioned
GZK energy losses or whether it is related to the maximum cosmic ray
acceleration energy at the sources.

A first comprehensive comparison of available data was performed by a
joint working group of Auger, TA, HiRes, and Yakutsk and is presented
in \cite{Dawson:2013wsa}. It is found that the energy spectra
determined by the Auger and TA observatories are consistent in
normalization and shape if the uncertainties in the energy scale -- at
that time quoted for each experiment to be about 20\,\% -- are taken
into account. This is a quite notable achievement and it demonstrates
how well the data of current observatories are understood. 

\begin{figure}[t,b]
\centerline{
\includegraphics[width=0.65\columnwidth]{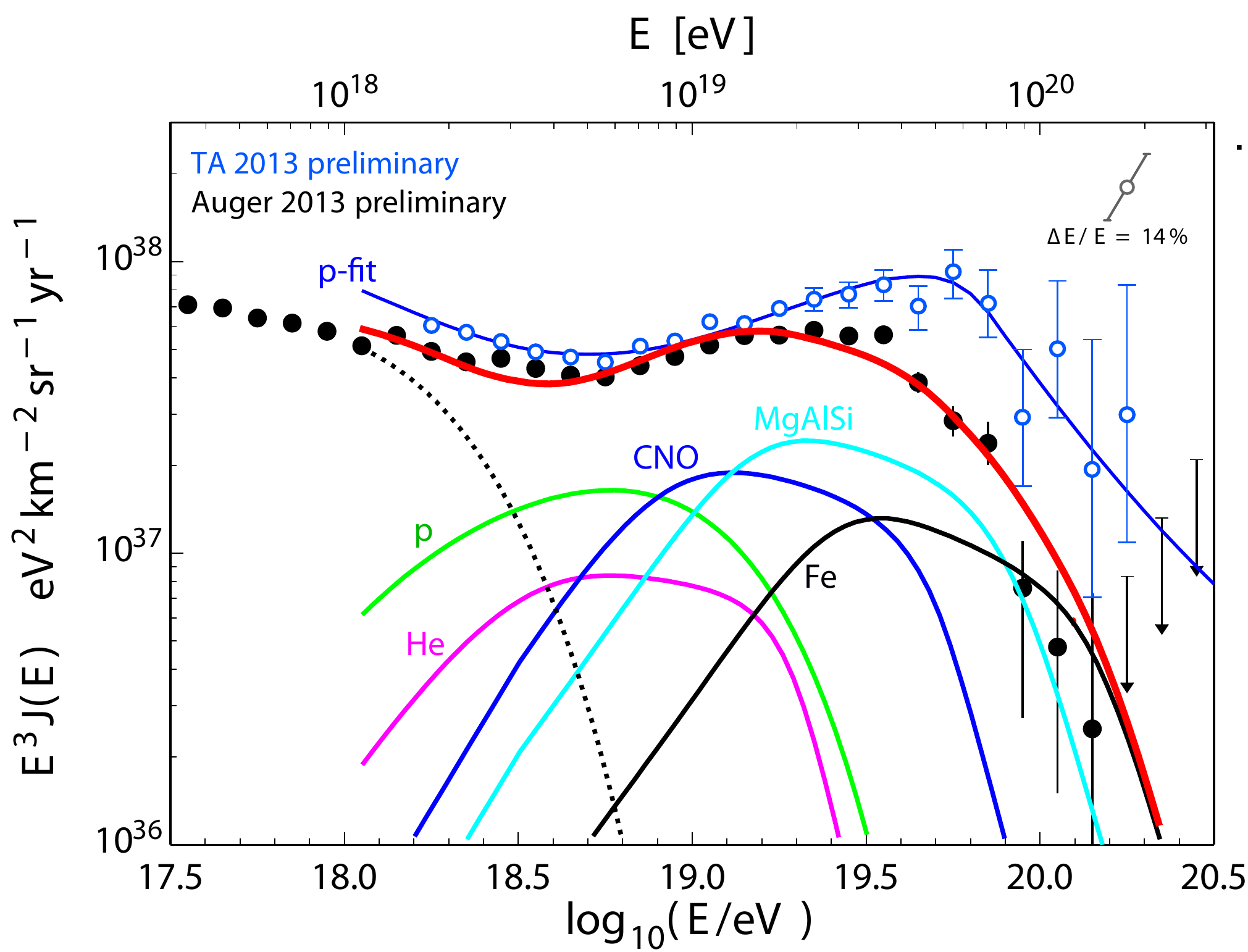}
}
\caption[xx]{Recent measurements of the flux of CRs at the highest
energies by the Auger and TA collaborations
\cite{Kido:2013hvr,Aab:2013ika}. The TA-data are fit to a model of
extragalactic proton sources, distributed cosmologically according to
$(1+z)^{4.4}$ and injecting a power-law distribution at the sources
according to $E^{-2.39}$ (blue line). The Auger data are compared to a
model assuming a maximum acceleration energy $E_{\rm
max}=10^{18.7}\,{\rm eV} \times Z$ with injection spectra $\gamma = 1$
and an enhanced galactic cosmic ray composition from
\cite{Aloisio:2013hya}. An additional galactic component is plotted as dotted black line. \label{fig:UHE-spectra}}
\end{figure}

Most recent updates of the cosmic ray energy spectra were presented at
the ICRC 2013 conference. Auger has reported an exposure of about
40\,000\,km$^2$\,sr\,yr in the zenith angle range up to $80^\circ$.
TA, due to the later start and its more than 4 times smaller area, has
collected about a $10^{\rm th}$ of the events. The TA collaboration
restricts the analysis to zenith angles below $45^\circ$ which can be
understood from the smaller vertical dimensions of the scintillator
slabs compared to the 1.2\,m height of the water tanks. Accounting
for recent precise measurements of the fluorescence yield
\cite{Ave:2012ifa} and taking advantage of a better estimate
of the invisible energy, a deeper understanding of the detector and
consequently improved event reconstruction, the Pierre Auger
Collaboration has recently updated their cosmic ray energy scale and
reduced its systematic uncertainties to 14\,\%
\cite{ThePierreAuger:2013eja}. The corresponding results of the two
experiments are presented in Fig.\,\ref{fig:UHE-spectra}. The energy
spectra of the two observatories clearly exhibit the ankle at $\sim
5\cdot10^{18}$\,eV and a flux suppression above $\sim
4\cdot10^{19}$\,eV, and are compared to simplified astrophysical
scenarios with parameters given in the figure caption. 

As can be seen from this comparison, the ankle occurs at an energy
which is compatible with the dip-model under the assumption of a pure
proton composition. Also, the flux suppression at the highest energies
is in accordance with the energy loss processes of the GZK-effect. In
the case of Auger, however, the suppression starts at lower energies
as compared to the propagation calculations unless the maximum energy
of sources is set to approx.\ $10^{20}$\,eV \cite{Aab:2013ika}. It is important to
realize that the suppression region of the spectrum can also be
described by assuming pure Fe-emission from the sources. In this case,
however, the ankle would require another component of cosmic rays to
contribute to the flux at lower energies. Another interpretation of
the suppression region has been presented in e.g.\
\cite{Aloisio:2009sj,Allard:2011aa,Biermann:2011wf,Taylor:2013gga}. In this group of
models, the flux suppression is primarily caused by the limiting
acceleration energy at the sources rather than by the GZK-effect. A
good description of the Auger all-particle energy spectrum is obtained for
$E_{\rm max,p} \simeq 10^{18.7}$\,eV with a mix of protons and heavier
nuclei being accelerated up to the same rigidity, so that their
maximum energy scales like $E_{\rm max,Z} \propto Z \times E_{\rm
max,p}$ (colored histograms in Fig.\,\ref{fig:UHE-spectra} \cite{Aloisio:2013hya}).
Obviously, the latter class of models (which also account for all relevant energy loss processes during propagation \cite{Kampert:2012fi}) leads to an increasingly heavier
composition towards the suppression region. We shall return to this
aspect in the next section. Another notable feature of such classes of
models is the requirement of injection spectra considerably harder
than those expected from Fermi acceleration. This was pointed out also
e.g.\ in Refs.\,\cite{Taylor:2013gga,Aloisio:2013hya,Gaisser:2013bla}.
However, as recently discussed in \cite{Mollerach:2013dza}, effects of
diffusion of high energy cosmic rays in turbulent extra-galactic
magnetic fields counteract the requirement of hard injection spectra
($\gamma < 2.0$) for a reasonable range of magnetic field strengths
and coherence lengths.

The different interpretations of the Auger and TA energy spectra
demonstrate the ambiguity left by the all-particle energy spectrum
and they underline the importance of understanding the absolute cosmic
ray energy scales to a high level of precision. While perfect
agreement is seen up to the ankle and beyond, one finds that the
flux-suppression in the Auger data not only starts at somewhat lower
energies, but also falls off more strongly than in TA data. This
difference -- despite being still compatible with the quoted
systematic uncertainties of TA and Auger of 20\,\% and 14\,\% --
deserves further attention.

\section{Mass Composition}
\label{sec:compos}

Obviously the all-particle energy spectrum by itself, despite the high
level of precision reached, does not allow one to conclude about the
origin of the spectral structures and thereby about the origin of
cosmic rays from the ankle to the highest energies. Additional key
information is obtained from the mass composition of cosmic rays.
Unfortunately, the measurement of primary masses is the most difficult
task in air shower physics as it relies on comparisons of data to EAS
simulations with the latter serving as reference
\cite{Kampert:2012mx,Engel:2011zz}. EAS simulations, however, are subject
to uncertainties mostly because hadronic interaction models need to be
employed at energy ranges much beyond those accessible to man-made
particle accelerators. Therefore, the advent of LHC data, particularly
those measured in the extreme forward region of the collisions, is of
great importance to cosmic ray and air shower physics and has been
awaited with great interest \cite{Kampert:2012mx}. Remarkably,
interaction models employed in air shower simulations provided a
somewhat better prediction of global observables (multiplicities,
$p_\perp$-distributions, forward and transverse energy flow, etc.)
than typical tunes of HEP models, such as PYTHIA or PHOJET
\cite{dEnterria:2011kw}. This revealed that the cosmic
ray community has taken great care in extrapolating models to the
highest energies. Moreover, as demonstrated e.g.\ in
\cite{Abreu:2012wt}, cosmic ray data provide important
information about particle physics at centre-of-mass energies ten or
more times higher than is accessible at LHC. The $pp$-inelastic cross
section extracted from data of the Pierre Auger Observatory supports
only a modest rise of the inelastic $pp$ cross section with energy
\cite{Abreu:2012wt}.

\begin{figure}
\centerline{
\includegraphics[width=.49\columnwidth]{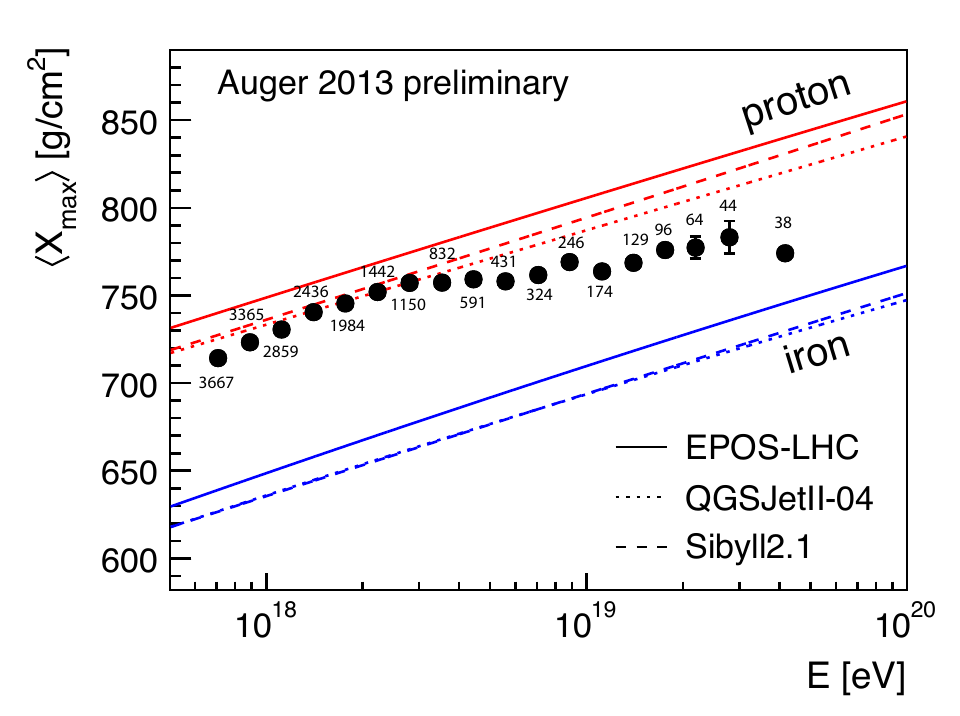}
\includegraphics[width=.49\columnwidth]{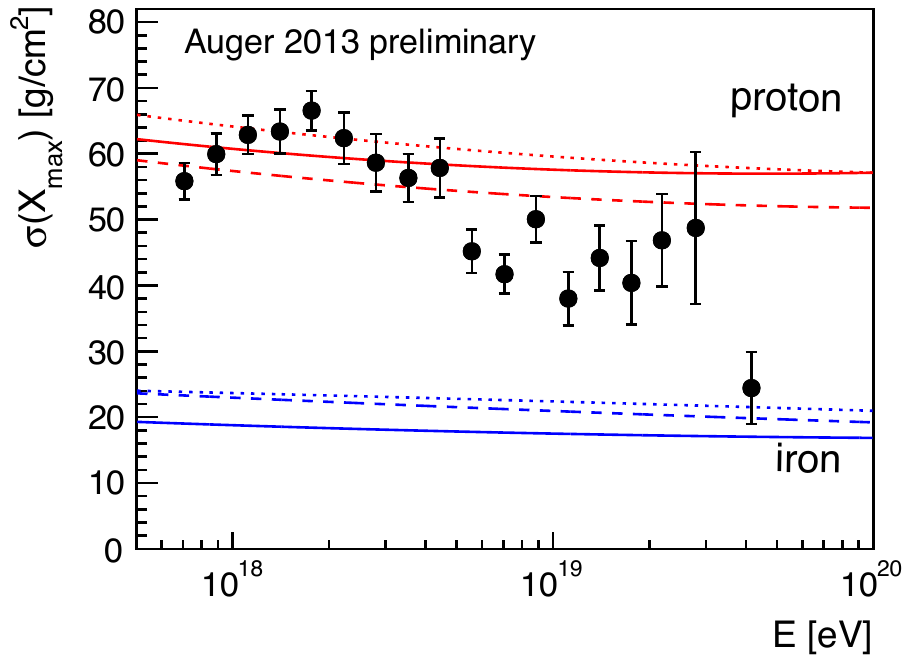}}
\centerline{
\includegraphics[width=.49\columnwidth]{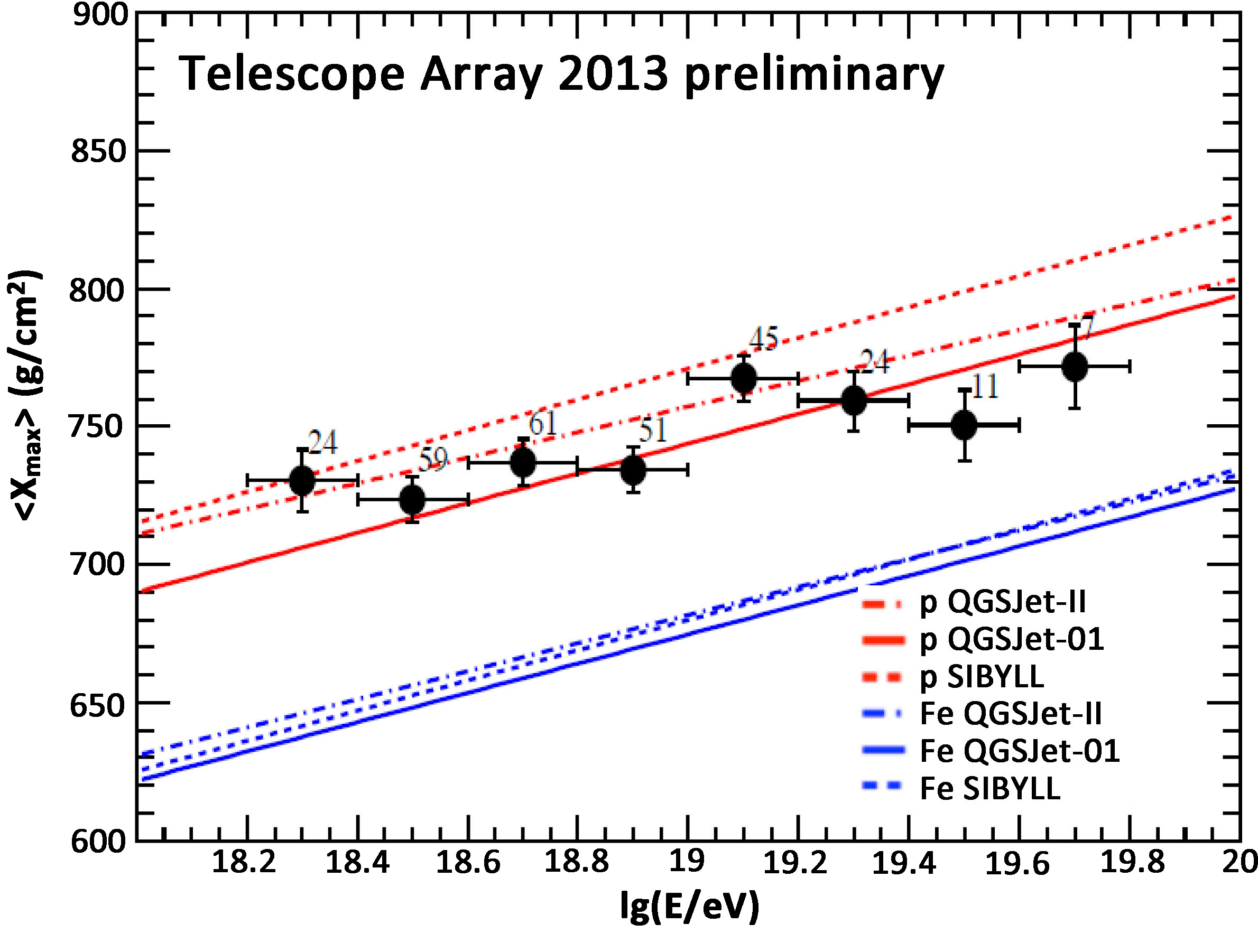}
\includegraphics[width=.49\columnwidth]{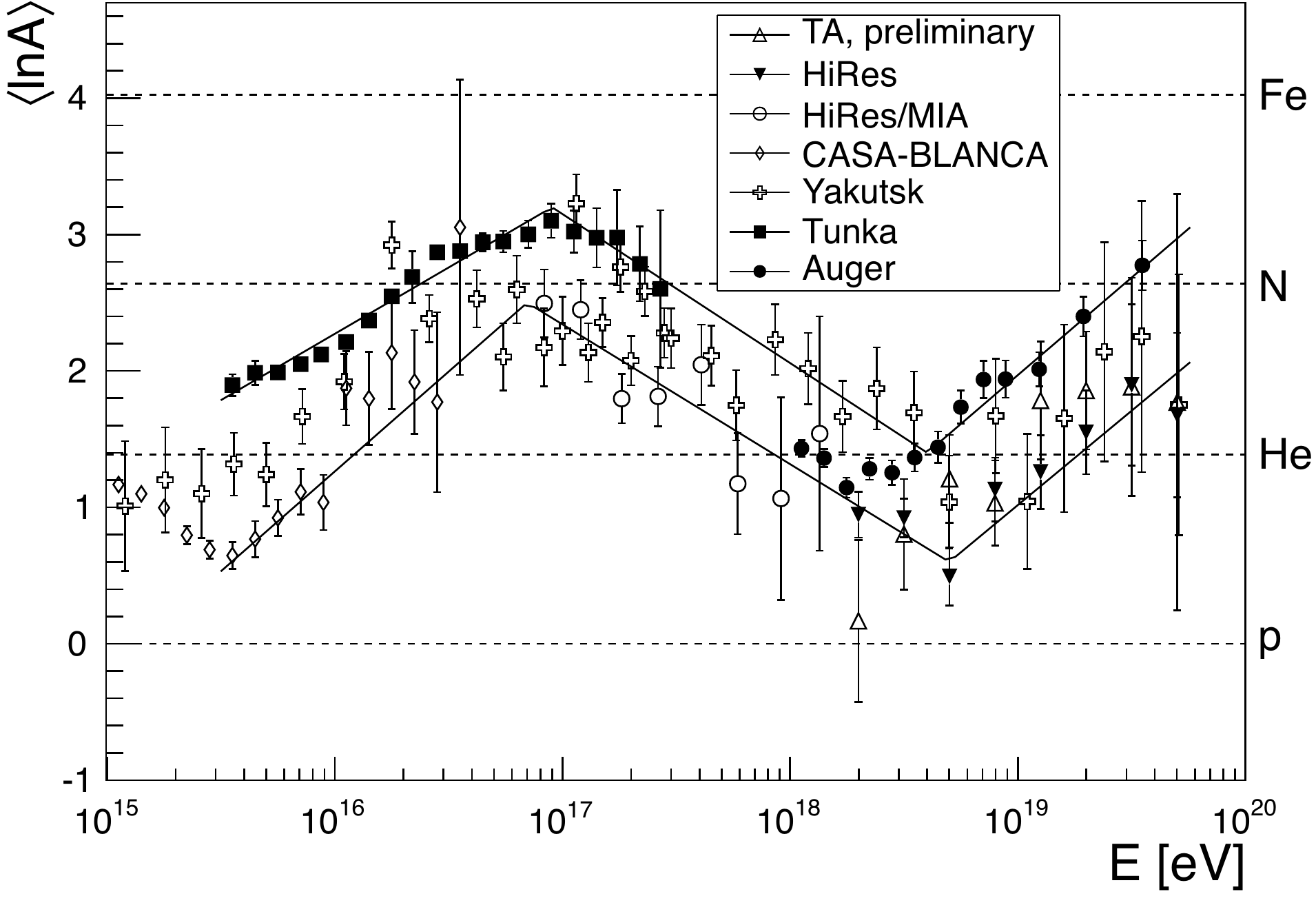}
}
\caption[xx]{Top: Evolution of $\langle X_{\rm max}\rangle$ and
$\sigma(X_{\rm max})$ with energy in data from the Pierre Auger
Observatory~\cite{Aab:2013ika}. Bottom left: $\langle X_{\rm
max}\rangle$ as a function of energy from TA~\cite{SagawaICRC2013}.
Bottom right: Average logarithmic mass of CRs as a function of energy
derived from $X_{\rm max}$ measurements with optical detectors for the
EPOS 1.99 interaction model. Lines are estimates of the experimental
systematics, i.e.\ upper and lower boundaries of the data presented
\cite{Kampert:2012mx}.\label{fig:compos}}
\end{figure}

A careful analysis of composition data from various experiments has
been performed and reviewed
in~\cite{Kampert:2012mx,Barcikowski:2013nfa}. Updated results from the
TA and Auger Observatories as well as a comparison of the two were
presented at the ICRC 2013 with exemplary results depicted in
Fig.\,\ref{fig:compos}. The data from the Pierre Auger Observatory
(Fig.\,\ref{fig:compos} top panel) suggest an increasingly heavy mass
composition above $4\cdot10^{18}$\,eV when compared to post-LHC
interaction models. The TA data are compatible with a proton dominated
composition at all energies (Fig.\,\ref{fig:compos} bottom left) but
have much larger statistical uncertainties and are compared to pre-LHC
interaction models which showed a larger scatter and mostly predicted
shallower showers. It is important to note that the datapoints and
model predictions of TA and Auger cannot be compared directly to each
other. This is because TA applies detector specific acceptance cuts to
data and Monte Carlo simulations while Auger applies fiducial volume
cuts aimed at selecting a bias free event sample. This is done by
using a high quality hybrid data set and applying fiducial volume cuts
based on the shower geometry that ensure that the viewable $X_{\rm
max}$ range for each shower is large enough to accommodate the full
$X_{\rm max}$ distribution \cite{Abraham:2010yv}. The price to be paid
for these so-called anti-bias cuts enabling a direct data-to-model
comparison is that it requires significantly
more statistics than the classical method of applying the same cuts to models and data. Because of this, it is presently not yet available in the TA data.
The event statistics surviving all cuts and entering in the $X_{\rm
max}$ energy bins of the Auger and TA data sample is specified in
Fig.\,\ref{fig:compos}. Because of these complications, both
collaborations have started to jointly investigate the origin of these
differences in $X_{\rm max}$ by injecting the measured composition
from the Pierre Auger Observatory into the TA Monte Carlo. The result
of that preliminary study shows that the proton- and Auger-like
composition cannot be discriminated from one another within the
presently available TA statistics \cite{HanlonICRC2013}. It will be
interesting to see this puzzle being solved in the near future both by
refined and improved reconstruction and analysis techniques, as well
as by collecting more data.

A (pre-ICRC 2013) compilation of composition data from various
experiments is depicted in Fig.\,\ref{fig:compos} (bottom right).
These data complement those of the energy spectrum in a remarkable
way. As can be seen, the breaks in the energy spectrum coincide with
the turning points of changes in the composition: the mean mass
becomes increasingly heavier above the knee, reaches a maximum near
the `iron-knee', another minimum at the ankle, before it starts
to modestly rise again towards the highest energies. Different
interaction models provide the same answer concerning changes in the
composition but differ by their absolute values of $\langle \ln A
\rangle$~\cite{Kampert:2012mx,Abreu:2013env}.

\begin{figure}
\centerline{
\includegraphics[width=.49\columnwidth]{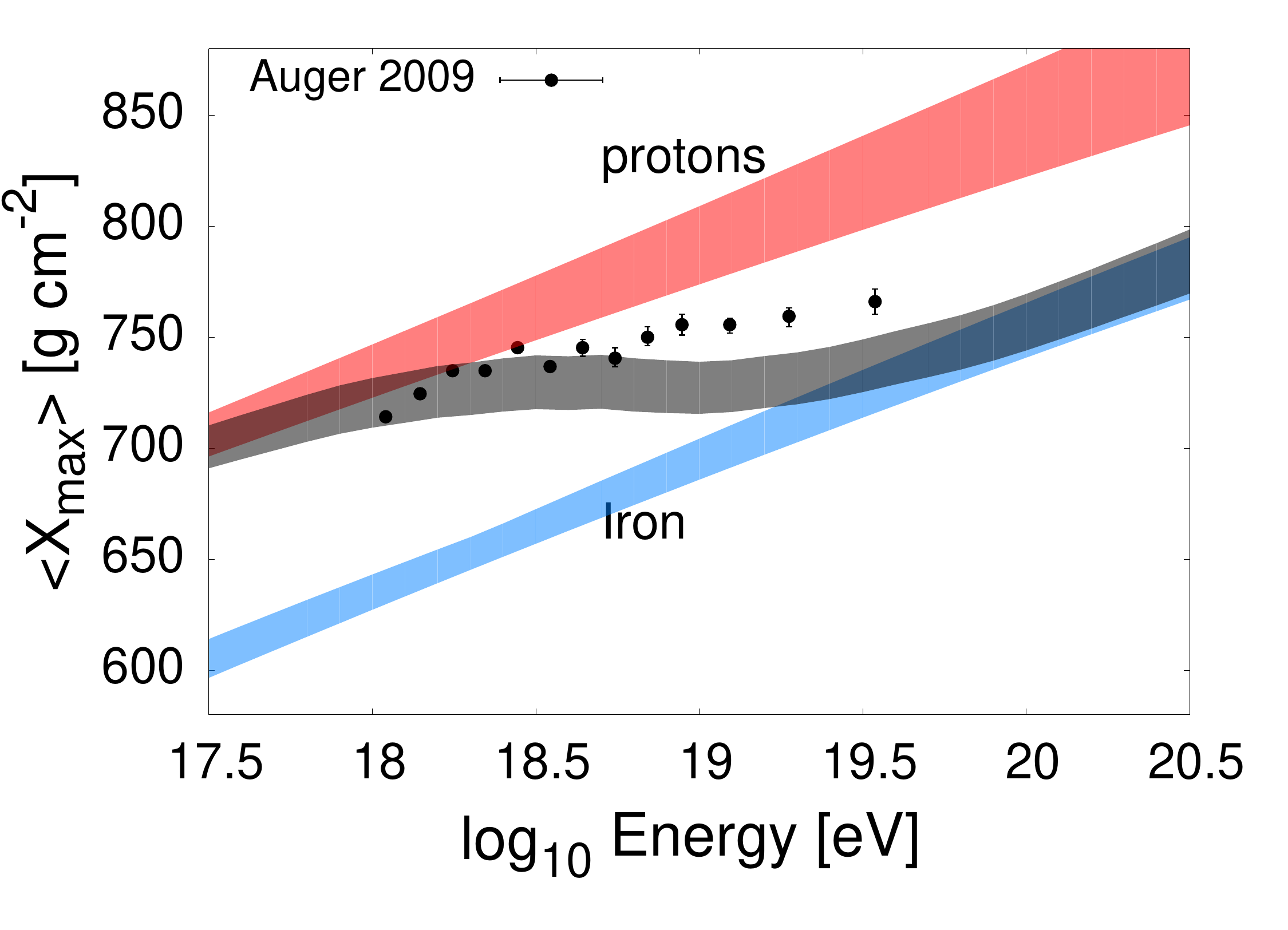}
\includegraphics[width=.49\columnwidth]{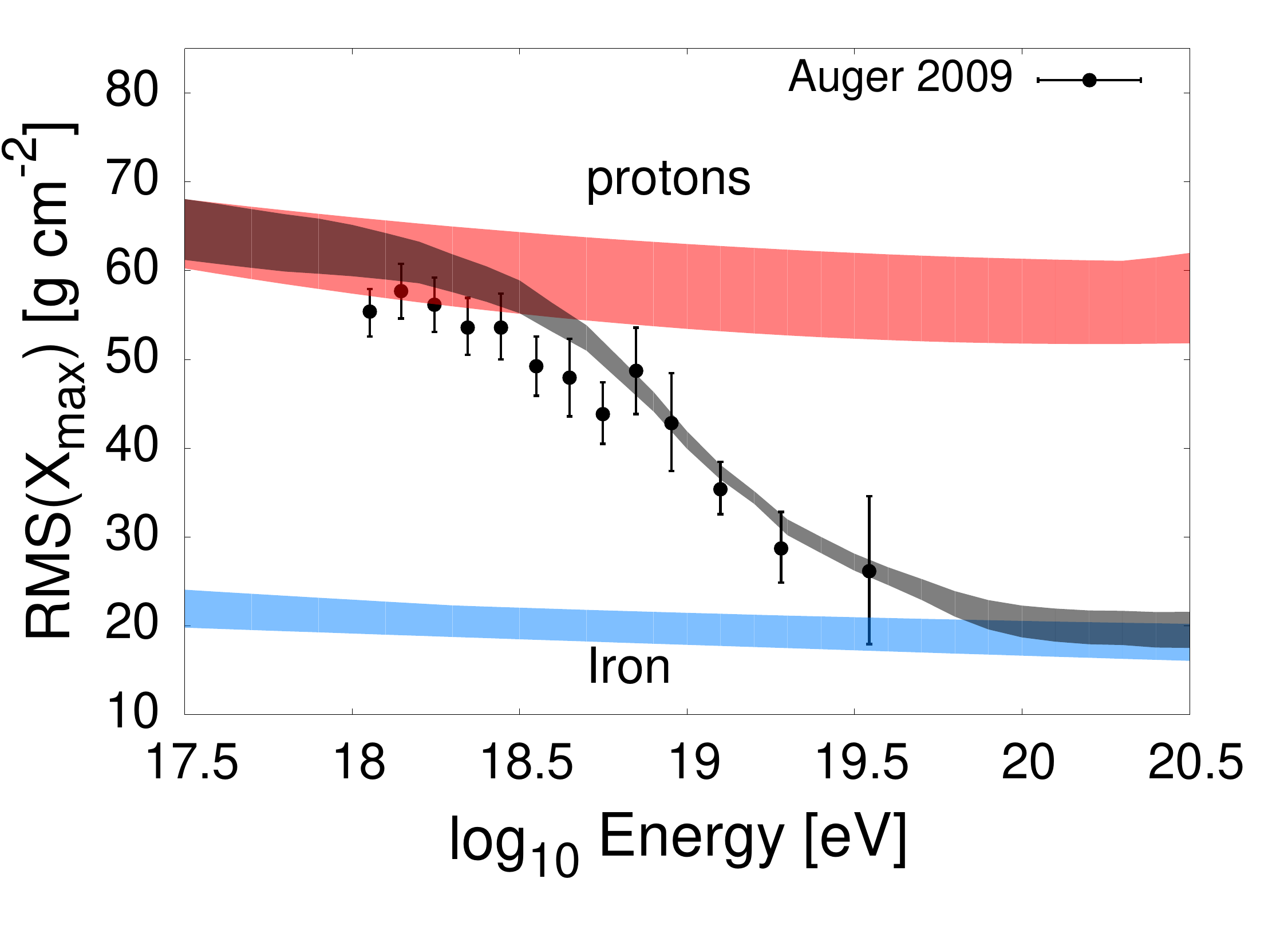}}
\caption[xx]{Example of $\langle X_{\rm max}\rangle$ and RMS$(X_{\rm
max})$ predicted by the archetypal maximum energy model of
\cite{Taylor:2013gga} in comparison to data from the Auger observatory.\label{fig:Taylor}}
\end{figure}

The interpretation of the all-particle energy spectrum in terms of the
exhaustion of sources rather than in terms of the GZK-effect,
discussed in the previous section (see histograms in
Fig.\,\ref{fig:UHE-spectra}), provides also a good description of the
evolution of $\langle X_{\rm max}\rangle$ and RMS$(X_{\rm max})$ with
energy, as seen by Auger. This is demonstrated exemplarily in
Fig.\,\ref{fig:Taylor} for the archetypal model from
Ref.\,\cite{Taylor:2013gga}. Similar results are reported e.g.\ in
Refs.~\cite{Hooper:2009fd,Aloisio:2013hya}.  

The mixture of light and intermediate/heavy primaries at the highest
energies predicted by the maximum-energy models may also explain the
low level of directional correlations to nearby AGN. Enhancements,
presently foreseen by the Pierre Auger Collaboration will address this
issue (see below). Moreover, improving the composition measurement in the ankle
region will be the key also to discriminate between different models
proposed to explain the transition from galactic to EG CRs. This has
been a prime motivation for the HEAT and TALE extensions of the Pierre
Auger and TA Observatories, respectively \cite{MathesICRC2011,ThomsonICRC2011}.
Clearly, the importance of measuring the composition up to the highest
energy cannot be overstated as it will be the key to answering the
question about the origin of the GZK-like flux suppression and the
transition from galactic- to extra-galactic cosmic rays discussed
above. 

\section{Anisotropies}
\label{sec:anisotropy}

\subsection{Data for anisotropy searches}

Further important information about the nature and origin of UHECR is
contained in the distribution of their arrival directions over
the sky. Unlike energies or primary mass, the arrival
directions of cosmic ray events are practically free from
systematic errors. 

Modern cosmic ray experiments are well suited for studying the
UHECR anisotropies at angular scales from about a degree up to
the largest scales corresponding to the whole sky. The bulk of
the arrival directions of UHECR events -- those measured by the
ground arrays -- have an angular resolution of about $\sim 1^\circ$
\cite{Bonifazi:2009ma,AbuZayyad:2012ru}. The angular resolution
may be up to an order of magnitude better for selected events
observed by the fluorescence detectors in the stereo or hybrid
modes \cite{Abraham:2009pm}, but the number of such events is
much smaller. Most of the anisotropy studies discussed in what
follows concerns data from the ground arrays. At $E>10^{19}$~eV
the total number of events accumulated to date exceeds $10^4$.

The ground arrays of both Auger and TA are fully efficient at
energies larger than $3\cdot10^{18}$\,eV \cite{Abraham:2010zz}
and $10^{19}$\,eV \cite{AbuZayyad:2012ru}, respectively. Above the
efficiency thresholds (and certainly above $10^{19}$~eV) the
integrated exposures of both experiments are very close to the
geometrical one \cite{AbuZayyad:2012hv}. This makes the
anisotropy study at high energies straightforward. Possible
(mild) deviations from the geometrical exposure have to be
studied and taken into account at energies below the efficiency
threshold. Together, Auger and TA cover the whole sky.

\subsection{Are anisotropies expected?}

Apart from the (unknown) distribution of sources over the sky, two main factors
that determine the UHECR anisotropy are deflections in cosmic magnetic fields
and attenuation due to the interactions with the radiation backgrounds. 

The extragalactic magnetic fields are known quite poorly. From measurements of
the Faraday rotations of extragalactic sources, they are usually assumed to
have a magnitude not exceeding $\lesssim 10^{-9}$~G \cite{Kronberg:1993vk} and
a correlation length up to $\sim 1$~Mpc. In such a field, a proton of
$10^{20}$~eV would be deflected by $\lesssim 2^\circ$ over a distance of
$50$~Mpc. Small deflections in the extragalactic fields are supported by
simulations \cite{Dolag:2004kp} which indicate that the extragalactic fields
are small everywhere except in galaxy clusters and filaments (see, however,
\cite{Sigl:2004yk
} and further discussion in 
\cite{Ryu:2008hi,Das:2008te,Ryu:2009pf}). The 
arguments based on the 
analysis of the gamma-ray propagation \cite{Essey:2010nd,Aharonian:2012fu} 
also point in this direction. 
An open, even though somewhat exotic, possibility is that
the Milky Way itself is embedded in a filament with relatively strong
magnetic fields, or that the galactic wind has magnetized the space around our
Galaxy \cite{Ahn:1999jd,Everett:2007dw}.

The Galactic magnetic field is known much better. Models of its regular
component have been constructed based on the existing measurements of the
Faraday rotations of extragalactic sources
\cite{Pshirkov:2011um,Jansson:2012rt}. This field would deflect a proton of
$10^{20}$~eV by about $2-4^\circ$ depending on the direction. The deflections
in the random component of the Galactic field were argued to be subdominant
\cite{Tinyakov:2004pw,Pshirkov:2013wka}. 

Energy losses of UHECR become important at energies in excess of
about $5\cdot 10^{19}$~eV (GZK-effect
\cite{Greisen:1966jv,Zatsepin66}). Although the mass composition
of UHECR is not known well, both protons and heavier nuclei are
subject to a similar attenuation and have a propagation horizon
of a few tens of Mpc at the highest energies.

As it is clear from the above numbers, if primary particles are predominantly
protons, one might expect to recover the distribution of sources over the sky,
with possibly bright spots of the size of a few degrees corresponding to
individual bright sources. On the other hand, if primary particles are heavier
nuclei, the flux distribution should be anisotropic in a manner similar (but
not identical) to the source distribution at the scale of a few tens of
degrees, but all the small-scale structure would be washed out. Note that
because of the small propagation distance, at the highest energies the sources are
expected to be distributed anisotropically due to the large-scale structure of
the Universe. 

None of these anisotropies are observed in the data. Below we summarize the
tests that have been performed, and discuss possible implications of the
results. 

\subsection{Searches for localized excesses of the UHECR flux} 

Two techniques are most commonly used to search for local excesses of
the UHECR flux. One is based on the two-point angular correlation
function (see, e.g., \cite{Tinyakov:2001ic} for the realization of
this method in the case of UHECR). This method is particularly useful
in cases when there are no very bright spots but rather many
excesses with a small amplitude and similar angular size. One then
expects an excess in correlations at the corresponding angular scale.
Both, Auger and TA data were examined in this way,
so far with negative results \cite{Abreu:2012zza,AbuZayyad:2012hv}.

Individual bright spots can be identified by looking for excesses in a
moving window of given angular size and estimating the background
either from Monte Carlo simulations or directly from the
data. The overall significance should be corrected for the effective
number of trials which is typically calculated by Monte Carlo
simulations. The Pierre Auger collaboration has performed this kind of
a blind search with window sizes of $5^\circ$ and $15^\circ$ in the
data set with energy $E>1$~EeV \cite{ThePierreAuger:2013eja}. No significant
excesses were found. In the TA data analogous searches were performed
in several energy bands around $1$~EeV with a search window of
$20^\circ$ \cite{KawataICRC2013-a} and a position-dependent window of
several degrees \cite{KawataICRC2013-b}. No significant deviation from
isotropy was found.

\begin{figure}
\centerline{\includegraphics[width=0.95\columnwidth]{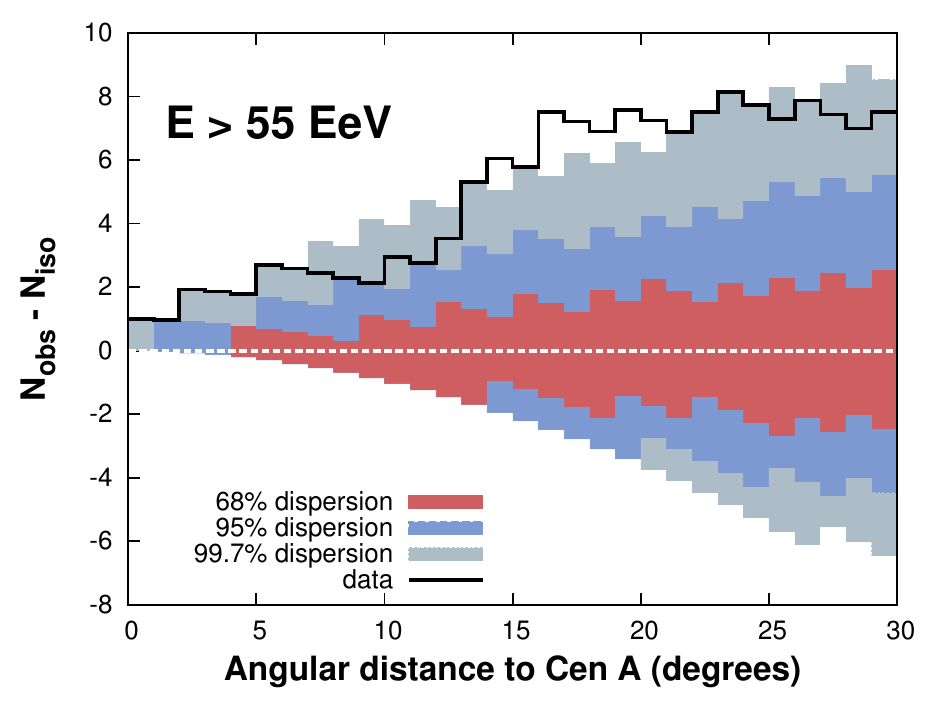}}
\caption{\label{auger-CenA} Cumulative number of events with $E >
55$~EeV as a function of angular distance from the direction of
Cen A \cite{Abreu:2011vm}. The isotropic background is
subtracted. The bands correspond to the dispersion expected for
an isotropic flux.}
\end{figure}

At high energies (around and above the cutoff in the spectrum)
the situation is more interesting. The Auger collaboration has
reported an excess of the UHECR events with $E>55$~EeV around the
direction towards the Centaurus supercluster at a distance of
about 60~Mpc and Centaurus A, a close AGN at a distance of about
3.5~Mpc. The largest excess was found for a circular region of
the angular size $18^\circ$. This region includes 10 out of 60
events above 55~EeV in the data set of this analysis, while 2.44
are expected from isotropy \cite{Abreu:2011vm}. At lower energies
no excess was found. The cumulative number of events (with the
background expectation subtracted) as a function of the angular
distance from the direction of Cen A is shown in
Fig.~\ref{auger-CenA} together with 1-, 2- and $3\sigma$ bands
representing fluctuations of the background.

\begin{figure}
\centerline{\includegraphics[width=0.95\columnwidth]{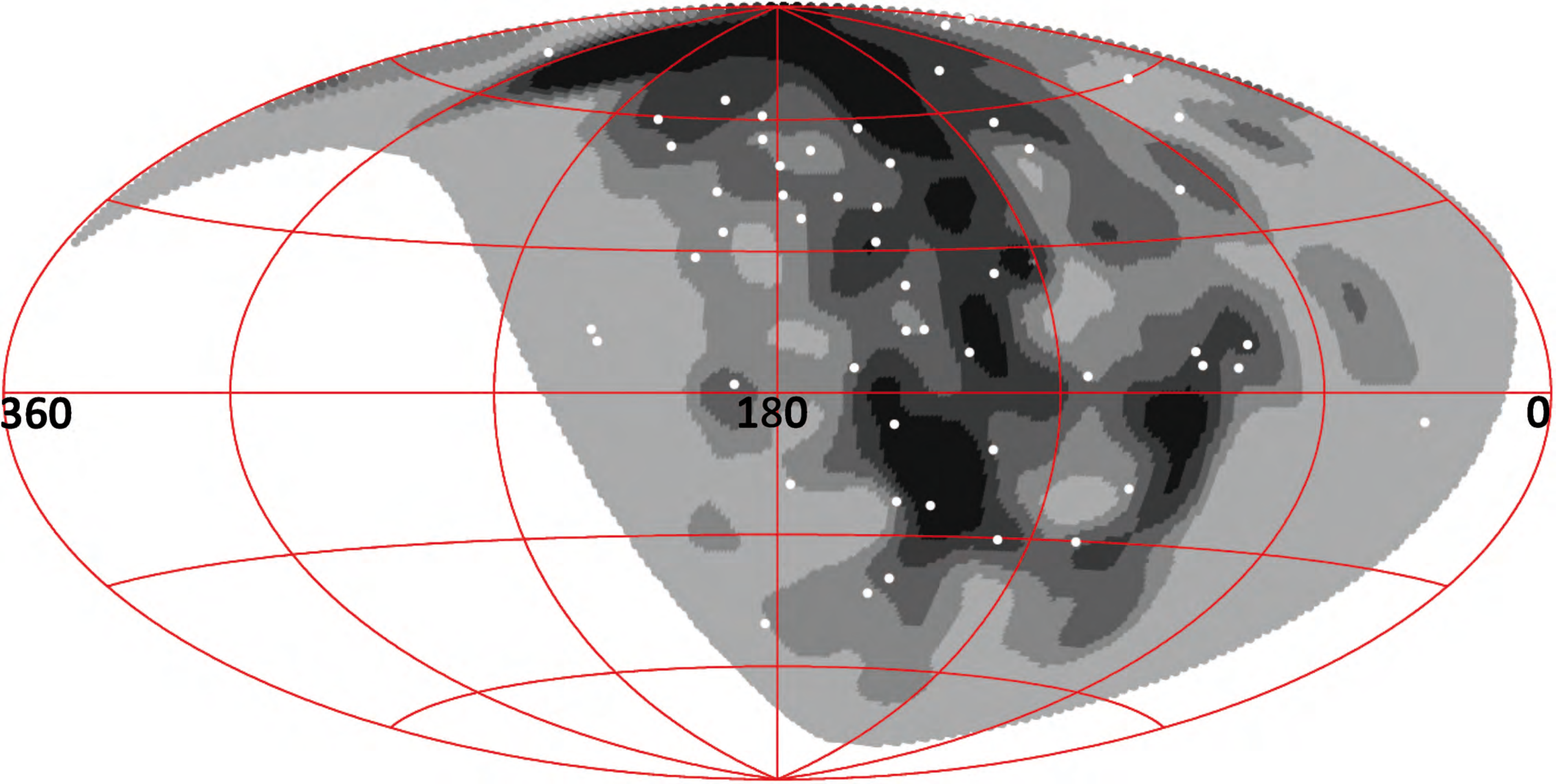}}
\caption{\label{TA-spot} The sky map of the TA events (white
dots) with $E>57$~EeV and the zenith angle cut $z<55^\circ$ in
the Galactic coordinates. The bands of grey represent the
expected UHECR flux assuming sources follow the matter
distribution in the local Universe, smeared with the angular
scale of $6^\circ$.}
\end{figure}

In the Northern sky, the TA collaboration has also observed some
deviation from isotropy in the data set with $E>57$~EeV at
similar angular scales \cite{Tinyakov-ICRC1033} in the direction about
$20^\circ$ from the Supergalactic plane, with no evident
astrophysical structures in the closer vicinity. The
corresponding sky map is shown in Fig.~\ref{TA-spot}. The
statistical significance of this ``hot spot'' has not been
reported. 

\subsection{Search for point sources} 

If the UHECR composition is light and the deflections are
dominated by the Galactic magnetic fields, or if the primary
particles are neutral, one might expect that at the highest energies
arrival directions of UHECR events roughly point back to their
sources. Because of the GZK cutoff, the UHECR propagation
distance of trans-GZK events, i.e.\ events exceeding the
GZK-threshold, is limited to 50-100~Mpc. The number of potential
sources of UHECR in this volume is limited, and one may expect
directional correlations between the position of candidate sources and the CR
event directions. This kind of analysis is complementary to the
one described above in the sense that it is optimized for the
situation when none of the sources is sufficiently bright to
produce a significant hot spot (cf.\ the discussion above).

\begin{figure}
\centerline{
\includegraphics[width=0.48\columnwidth]{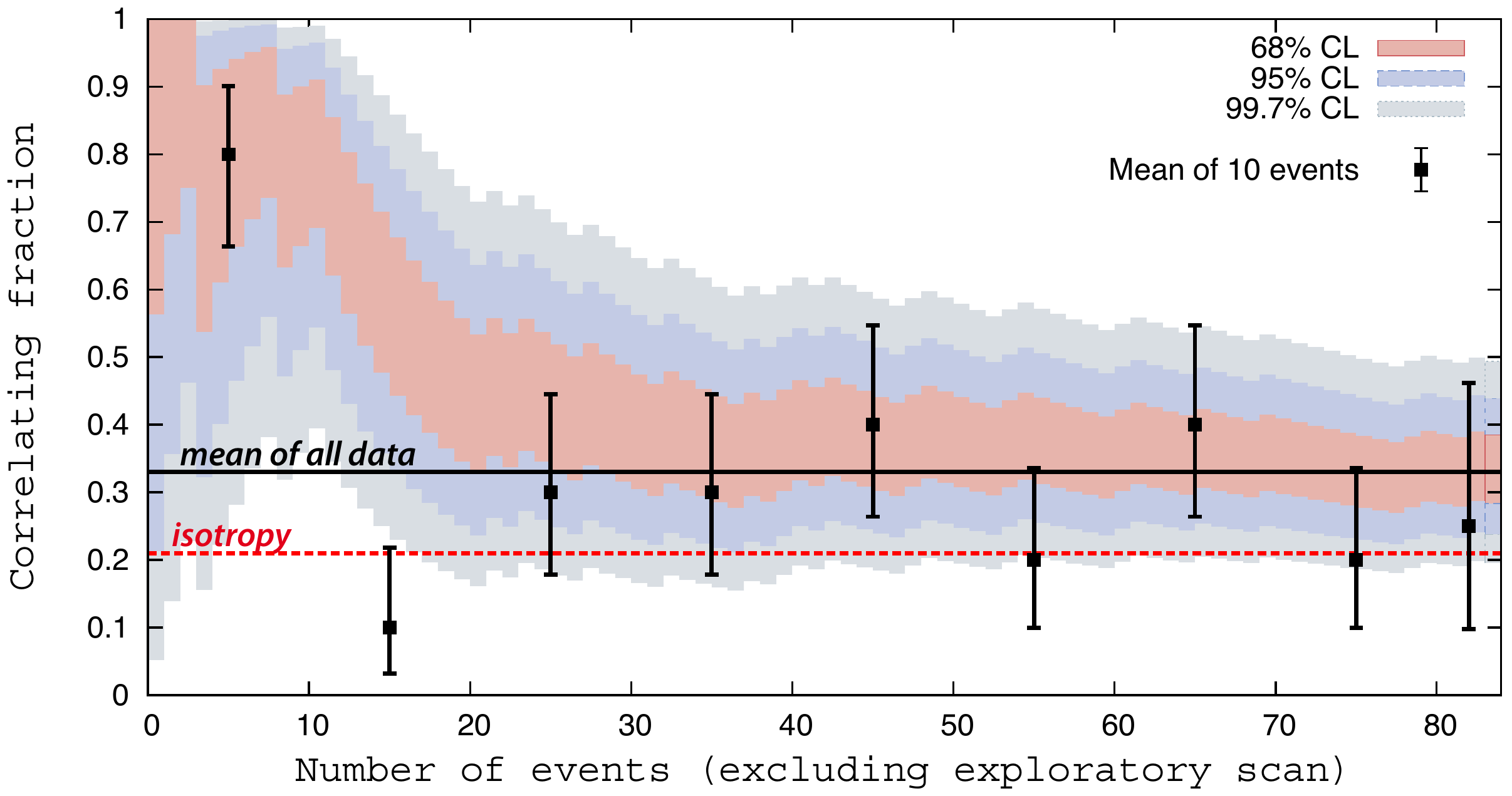}
\includegraphics[width=0.48\columnwidth]{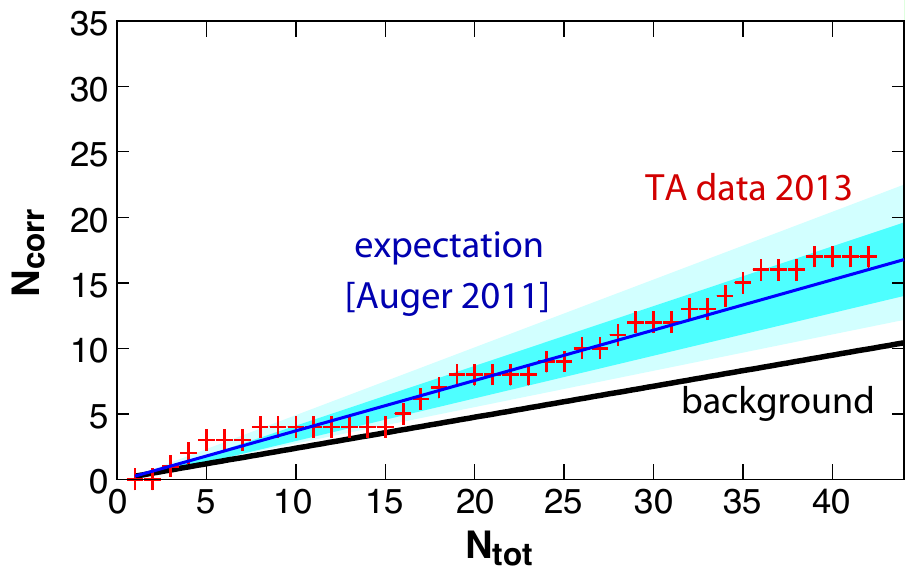}}
\caption{\label{auger-AGN}
Left: The most likely value of the degree of correlation $p_{data} = k/N$ is plotted as a function of the
total number of time-ordered events (excluding the data used for the parameter scan)
\cite{Kampert:2012vh}. The 68\%, 95\% and 99.7\% confidence level
intervals around the most likely value are shaded. The horizontal
red dashed line shows the isotropic value $p_{iso} = 0.21$ and the
full black line the current estimate of the signal $p_{data} = 0.33 \pm
0.05$. The black symbols show the correlation fraction in
independent bins with 10 consecutive events.
Right: Number of correlating events from TA (red
crosses) \cite{Tinyakov-ICRC1033} as a function of the total number
of events.  The black line shows the expected number of random
coincidences assuming a uniform background. The latest data
correspond to 17 correlating events out of 42. The shaded area shows
the expectation (1- and 2$\sigma$ bands) based on the degree of
correlation measured by Auger \cite{Kampert:2012vh}.
}
\end{figure}

The Auger collaboration has studied the correlation of the
highest energy events above $55$~EeV with the nearby Active
Galactic Nuclei (AGNs) from the V\'eron-Cetty and V\'eron catalog
(VCV) \cite{VeronCetty:2003xx}. The parameters of the correlation
(the energy threshold at $55$~EeV, the maximum distance in the
catalog of $75$~Mpc and the maximum opening angle of $3.1^\circ$)
were fixed from the exploratory scans in the independent data set
\cite{Abraham:2007bb,Abreu:2010zzj}. The latest results of this study
\cite{Kampert:2012vh} is presented in Fig.~\ref{auger-AGN} (left) which
shows the most likely fraction of correlating events plotted as a
function of the total number of events, together with the
$1,2,3-\sigma$ bands which allow one to see how far the observed
number of correlated events deviates from the expectation
assuming an isotropic background. One can see that while in the
early part of the data there was a substantial deviation from
isotropy, with the accumulation of events the correlation
strength has decreased to $33 \pm 5$\% compared to $21$\%
expected from isotropy. The statistical significance for a
departure from isotropy has over this period remained almost
constant at a level between $2$ and $3\sigma$.

Correlation with the same set of AGN and with the parameters
fixed at the values set by the Auger collaboration analysis has
been studied by the HiRes collaboration \cite{Abbasi:2008md}
with a negative result, and by the TA collaboration
\cite{AbuZayyad:2012hv}. The most recent update of the TA
analysis is presented in Fig.~\ref{auger-AGN} (right) which shows the number
of correlating events as a function of the total number of
events. There is a slight excess of correlating events over the
expected background, compatible with both the background and with
the latest update on the AGN correlations from Auger. The expectation from the latest Auger data \cite{Kampert:2012vh} is depicted by the 1- and 2$\sigma$-bands which demonstrates an excellent agreement of the two data sets. The combined probability to observe such a correlation from an isotropic distribution is below $p=10^{-3}$, still too large to draw any firm conclusions.

\subsection{Harmonic analysis} 

A standard tool in search for medium and large scale anisotropy
searches is the harmonic analysis. In the case of UHECR, the
application of this method is limited by the incomplete sky
coverage of presently existing observatories which cover either
the southern (in case of Auger) or northern (in case of TA) part
of the sky. For this reason, not all components of the low
multipoles can be extracted unambiguously from the data of a
single experiment. For instance, because of the (approximate)
azimuthal symmetry of the exposure function, only the
$(xy)$-components of the dipole (in equatorial coordinates) can
be obtained in a straightforward way by a single experiment.

\begin{figure}
\centerline{
\includegraphics[width=0.45\columnwidth]{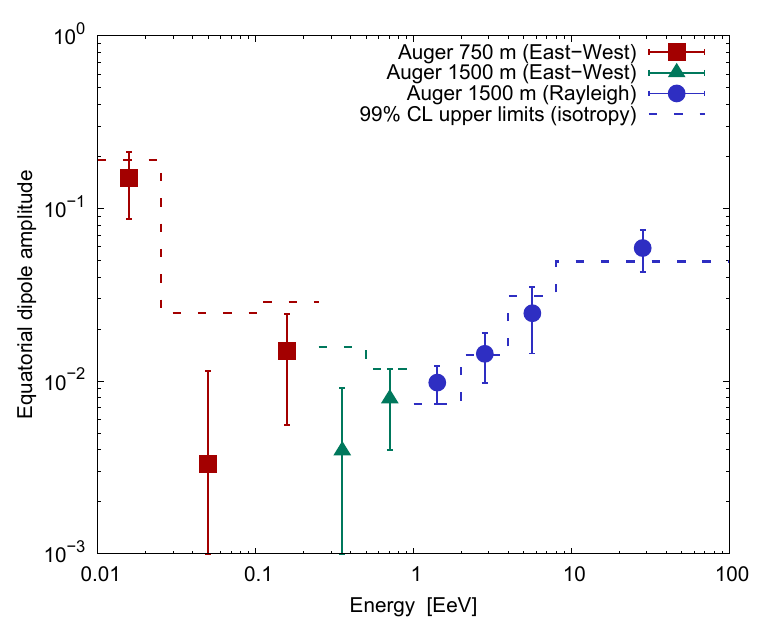}
\includegraphics[width=0.45\columnwidth]{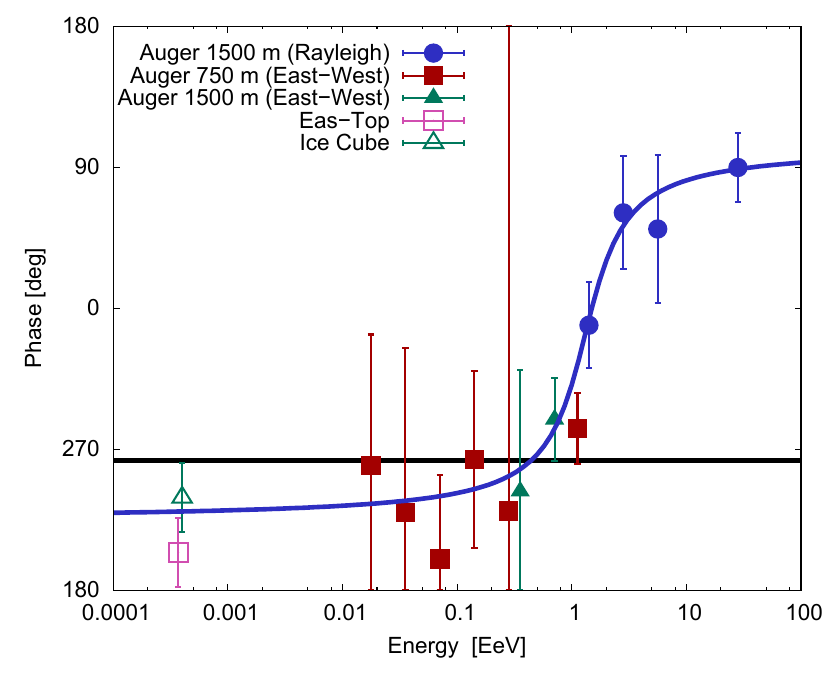}
}
\caption{\label{dipole} {\em Left panel:} Equatorial dipole
amplitude as a function of energy. The results of the modified
Rayleigh analysis are shown with black circles and blue triangles
corresponds to the analysis with East-West method. Red squares
correspond to data from the infill array using the East-West
method.  The dashed lines are the 99\% CL upper values of the
amplitude that could result from fluctuations of an isotropic
distribution. {\em Right panel:} Phase of the first harmonic as a
function of energy. The horizontal black line corresponds to the value
$\phi =263^\circ$ roughly coincident with the azimuthal direction
to the Galactic center. The continuous blue curve is the fit to an
empirical formula performed in \cite{Abreu:2011ve}.  }
\end{figure}

Results of a search for the equatorial dipole have been reported
by the Pierre Auger collaboration
\cite{ThePierreAuger:2013eja,Abreu:2012ybu}. Fig.~\ref{dipole}
(left panel) shows the measurement of the dipole amplitude as a
function of energy. Different analysis techniques have been used
in different energy bins as indicated in the plot. The measured
amplitude of the dipole is consistent with expectations from the
isotropic background. It is interesting to note, however, that
the dipole amplitude is not the most sensitive observable
\cite{Abreu:2012ybu} because of the energy binning and related
loss in statistics. Even when the dipole amplitude is not
sufficiently large to be detected, its phase may show regular
behavior with energy, which would be an indication for a non-zero
dipole. The right panel of Fig.~\ref{dipole} shows the phase of
the dipole as a function of the energy. One can observe that the
values of the phase are correlated in adjacent energy bins, and
the phase behavior with energy is consistent with a continuous
curve. This may indicate the presence of a non-zero dipole in the
Auger data whose amplitude is just below the detection threshold.

The problem of the incomplete sky coverage may be resolved by
combining the data of the two observatories. This is not a
straightforward procedure because of the uncertainty in the relative
flux calibration resulting mainly from possible differences in the
energy scales of experiments. The difficulty, however, may be
overcome, and the corresponding analysis is presently underway
\cite{Auger-TA-LSA:2014} with the first all-sky UHECR intensity presented at the ICRC 2013 with no significant under/overdensities found, yet \cite{Array:2013dra}. 

\subsection{Large-scale anisotropy}

If the deflections of UHECR do not exceed $10-20^\circ$, as 
in the case of (predominantly) proton composition and small
extragalactic magnetic fields, one should expect a correlation of
UHECR arrival directions with the local large-scale structures
(LSS). The largest correlations are expected at or above the
GZK-threshold energy, because in this energy range the propagation
distance is limited to $50-100$~Mpc and the contributions of the
local structures is enhanced. With enough statistics, by checking
such a correlation one may either discover it, or put a lower
limit on the UHECR deflections. With some assumptions about
cosmic magnetic fields, this information may also help to
understand the UHECR composition.

\begin{figure}
\centerline{\includegraphics[width=0.95\columnwidth]{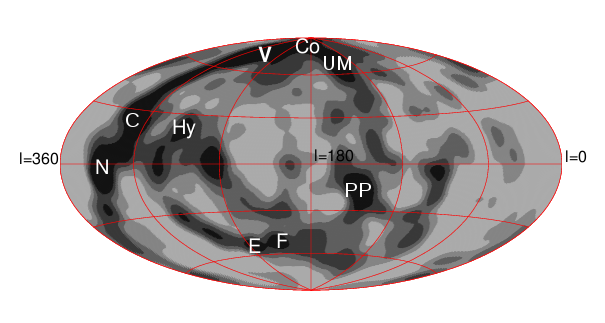}}
\caption{\label{TA-LSSmap} Sky map of UHECR flux expected in a model
where the sources follow the matter distribution, in Galactic
coordinates. Darker regions correspond to larger flux. Each band
integrates to 1/5 of the total flux. Letters indicate nearby matter
structures: C: Centaurus supercluster (60 Mpc); Co: Coma cluster (90
Mpc); E: Eridanus cluster (30 Mpc); F: Fornax cluster (20 Mpc); Hy:
Hydra supercluster (50 Mpc); N: Norma supercluster (65 Mpc); PI:
Pavo-Indus supercluster (70 Mpc); PP: Perseus-Pisces supercluster (70
Mpc); Ursa Major North group (20 Mpc) South group (20 Mpc); V: Virgo
cluster (20 Mpc).  }
\end{figure}

The distribution of the UHECR flux expected in a generic model where
sources trace the distribution of matter in the nearby Universe was
calculated, e.g., in Ref.~\cite{Koers:2008ba}. An improved version of
this map obtained using a larger catalog of galaxies is presented in
Fig.~\ref{TA-LSSmap}. This map was calculated assuming the UHECR are
protons of energy 57~EeV, and smeared over an angular scale of
$6^\circ$.

\begin{figure}
\includegraphics[width=0.49\columnwidth]{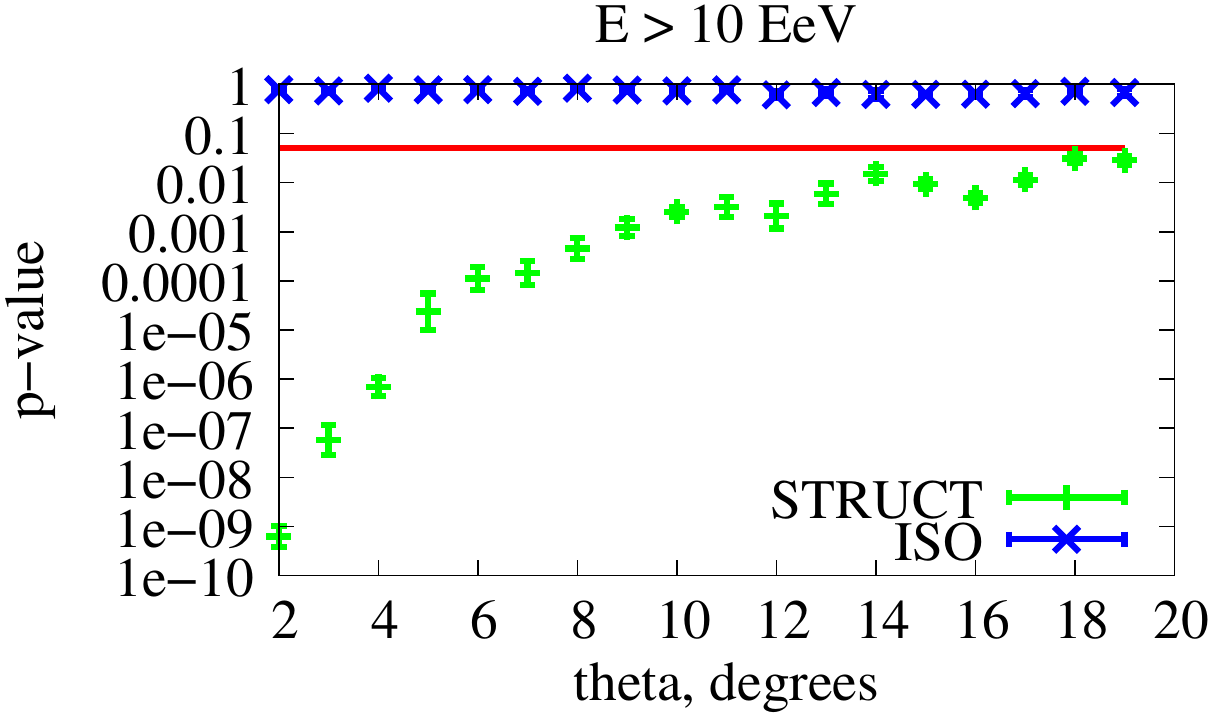}
\includegraphics[width=0.49\columnwidth]{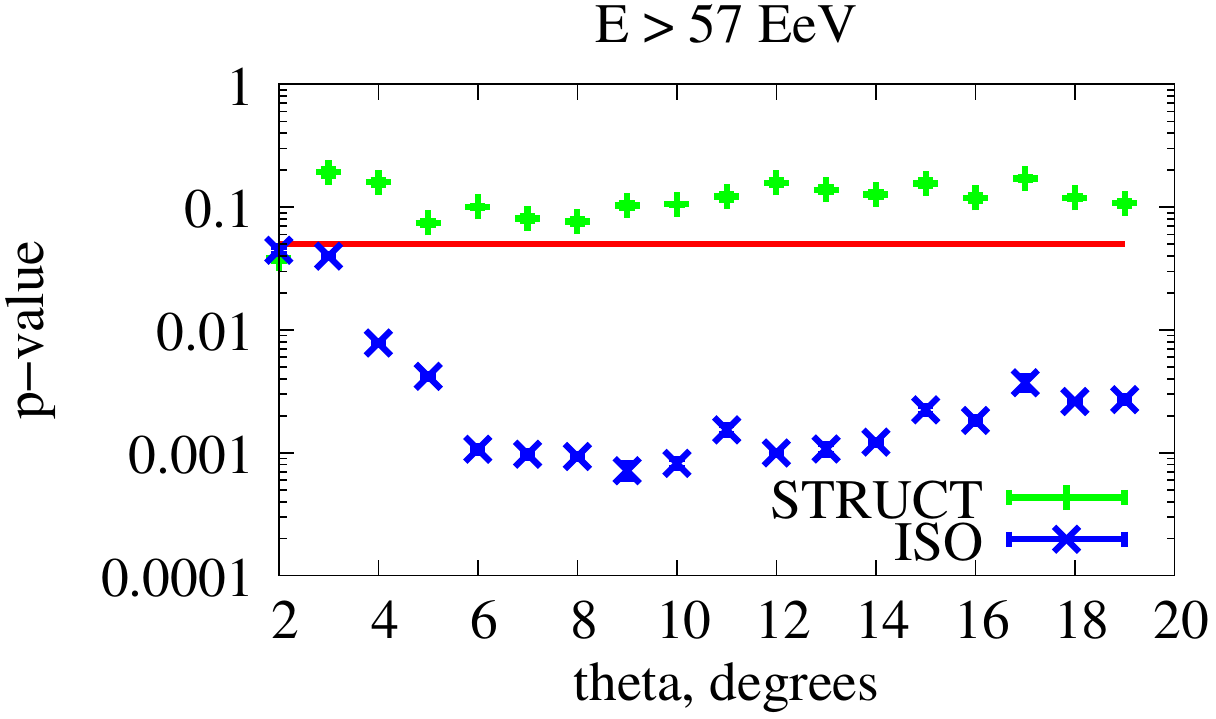}
\caption{\label{TA-LSStests} The results of a statistical test for
  correlations between the LSS at different smearing angles $\theta$
  and the TA data with $E>10$~EeV (left panel) and $E>57$~EeV. Green
  points represent $p$-values corresponding to the LSS model, blue
  points -- to the isotropic distribution.}
\end{figure}

The expected flux map may be compared to the actual UHECR distribution
by making use of an appropriate statistical test (see, e.g.,
\cite{Koers:2008ba}). The results of the analysis using the latest TA
data set are shown in Fig.~\ref{TA-LSStests} for two datasets with
$E>10$~EeV and $E>57$~EeV. One can see that at low energies
$E>10$~EeV the data are compatible with isotropy and incompatible with
the LSS model for all but largest smearing angles. At high energies,
on the contrary, the data are compatible with the structure and not
compatible with isotropy (the latter may be another manifestation of the ``hot spot'' discussed above). 

A similar analysis has been performed using the first 69 publicly
released Auger data \cite{Oikonomou:2012ef} with energies $E>55$~EeV.
It was found that the correlation of the Auger events with the LSS
prediction is larger than would be in the isotropic model, but smaller
than in the model where the UHECR sources follow the matter
distribution in the Universe. 

\subsection{Other searches} 

If galactic TeV gamma-rays originate from energetic protons suffering
pion-production interactions with ambient photons, protons, or
nuclei, one should expect that neutrons are also produced. At energies
higher than $10^{18}$~eV neutrons can reach us from large parts of the galaxy before they
decay ($\tau_{\rm n} = 9.2\,{\rm kpc}\times E$/EeV). Since neutrons are not deflected by the magnetic fields, they
should point back to their sources. 

The Pierre Auger Collaboration has performed a dedicated search for
Galactic sources of neutrons \cite{Auger:2012yc}. Several classes of
sources were considered, such as H.E.S.S.\ TeV sources, several
classes of pulsars, microquasars, and magnetars. These sources were
stacked in their respective classes. The search window was set to the
angular resolution of the detector. In addition to these sources, the
Galactic plane and the Galactic Center were considered as possible
sources. The advantage of this analysis over the blind search is that
the penalty for trials is substantially reduced. No statistically
significant excess was detected in any of the catalogs, including the
Galactic plane and the Galactic Center. 

In a related analysis \cite{KuempelICRC2013}, a search for point sources of EeV photons was performed. With no photon point source being
detected, upper limits on the photon flux have been derived for every
direction within the Auger exposure map. None exceeds an energy flux of
0.25~eV\,cm$^{-2}$\,s$^{-1}$ in any part of the sky assuming a photon
flux following $1/E^2$. These limits are of considerable astrophysical
interest, because the energy flux in TeV gamma rays exceeds
1~eV\,cm$^{-2}$\,s$^{-1}$ for some Galactic sources with a
differential spectral index of $E^{-2}$ \cite{Hinton:2009zz}.

\section{Conclusions and Outlook}
\label{sec:summary}

To summarize, the new generation of experiments -- the Pierre Auger
Observatory and the Telescope Array -- have been constructed and
operated in the last decade. Both experiments proved the advantage of
the hybrid detector design where the fluorescence telescopes are
combined with the ground array of detectors. The former are used for
calorimetric energy measurements and calibration of the ground array
energy scale, while the ground array takes advantage of its 100\% duty
cycle to accumulate large statistics. As a result, the uncertainty in
the energy estimate has been reduced to much below 20\%, and more than
10-fold increase in statistics has been achieved.

This has lead to a number of important advances. First, the features
in the UHECR energy spectrum -- the ankle and the suppression at the
highest energies -- have been established beyond doubt. The spectral
slopes before and after the ankle have been measured to the second
digit and agree between the two experiments. The positions of the
ankle also agree within the quoted errors, and are compatible with the
existing model(s). The parameters of the break at the highest energies are
known less accurately. There seems to be some discrepancy concerning
the shape of the spectrum around the break, however more statistics is
needed for a firm conclusion. The position of the break is compatible
with the GZK cutoff for protons, but other explanations are also
possible. 

The substantial increase in statistics allowed one to put stringent
constraints on the previously claimed deviations of the arrival directions
from the isotropic distribution. This concerns the clustering of the UHECR
events, as well as their correlations with different classes of putative
sources. Unfortunately, no significant deviation from isotropy has been
confirmed, yet. 

As far as the mass composition of UHECR is concerned, the situation is
less definite, and a consistent picture has not yet emerged.
While the Pierre Auger Observatory sees a change of the composition
towards a heavier one at the highest energies, the TA observes no such
a trend and is compatible with a pure proton composition. This
difference in the data has profound consequences: The Auger data
suggest that we see the maximum energy of sources, similarly to what
is observed at the knee in the cosmic ray spectrum, while the TA data
suggest we observe the GZK-effect. Seeing the GZK-effect would
naturally allow to interprete the ankle in terms of $e^+e^-$-pair
production losses in the CMB while the maximum energy scenario relates
the ankle to the transition of galactic to extragalactic cosmic rays.
The hard injection spectra required by the maximum energy model would
either call for non-standard acceleration processes or require a
contribution of nearby sources to the all-particle flux. Moreover, the
different compositions in the GZK- and maximum-energy scenario will
affect the level of anisotropies expected to be seen in the data. As
already mentioned, a pure proton composition up to the highest energies starts
to conflict with the highly isotropic UHECR-sky, unless extremely strong
galactic and extragalactic magnetics fields are assumed.

Thus, despite the major advances, a number of key questions remain
open: (i) a more accurate absolute energy calibration is needed to
clarify the physical interpretation of the ankle and the high-energy
break in the spectrum; (ii) the apparent differences in the observed
mass composition at highest energies need to be understood; a more
accurate modeling of air showers may be required for that in addition
to a better understanding of systematic biases; (iii) the apparent
absence of anisotropies, especially at the highest energies, has to be
reconciled with the mass composition and our knowledge of the cosmic
magnetic fields and the existing source models. 

An important lesson from the existing picture is that the above open
problems are closely interrelated. It is not inconceivable that a
breakthrough in one of these questions will lead to the understanding
of the others and finally to the emergence of a consistent picture of
UHECR. The next advance in the experimental techniques, presently
prepared by both collaborations, is therefore likely to be the last
crucial step in our understanding of the nature and origin of these
highest-energy particles ever observed in Nature.

\section*{Acknowledgements}
We gratefully acknowledge stimulating discussions with out colleagues
in the TA and Pierre Auger Collaborations. KHK acknowledges financial
support by the German Ministry of Research and Education (Grants
05A11PX1 and 05A11PXA) and by the Helmholtz Alliance for Astroparticle
Physics (HAP). PT acknowledges the support of the IISN project No.
4.4502.13, the RFBR grant 13-02-12175-ofi-m and Belgian Science Policy
under IUAP VII/37.

\end{document}